\documentclass[aps,pra,twocolumn,showpacs,floatfix]{revtex4}
\usepackage{epsfig}
\usepackage{amsthm,amsmath}
\usepackage{bm}
\usepackage{bbold}

\begin{document}
\title{Relativistic equation-of-motion coupled-cluster method
for the ionization problem: Application to molecules}
\author{Himadri Pathak$^{1}$\footnote{h.pathak@ncl.res.in.\\first two authors contributed equally.}, Sudip Sasmal$^{1}$,
 Malaya K. Nayak$^2$, Nayana Vaval$^1$ and Sourav Pal$^1$}
\affiliation{$^1$ Electronic Structure Theory Group,
 Physical Chemistry Division,
 CSIR-National Chemical Laboratory,
 Pune, 411008, India}
\affiliation{$^2$ Bhabha Atomic Research Centre,
 Trombay, Mumbai - 400085, India}
\begin{abstract}
We report the implementation of 4-component spinor relativistic equation-of-motion coupled-cluster
method within the single- and double- excitation approximation to calculate ionization potential (EOM-CCSD) of
molecules. We have applied this method to calculate vertical ionization potentials of the molecules,
XH(X=F, Cl, Br, I) along with Cl$_2$ and Br$_2$ in their closed-shell configuration.
We have also presented intermediate results using 2-nd order many-body perturbation theory
level in the EOM framework (EOM-MBPT(2)) to understand the role of electron correlation.
All the calculated values are compared with the available experimental results.
Our results are found to be in well agreement with the sophisticated experiments and 
relative deviation of less than 1$\%$ achieved for all the considered systems.
\pacs{31.15.bw, 31.15.vn, 33.15.Ry}
\end{abstract}
\maketitle
\section{Introduction}
It is always a hard task for the theoretical physicists to come up with a method
that can depict the atomic and molecular spectroscopic properties very precisely.
The effects of electron correlations as well as the relativistic effects
have to be taken care simultaneously as they are intertwined \cite{1}. The Dirac-Hartree-Fock
method in its four-component formalism is the best possible way to include the effects of the relativity
within a single determinantal description. On the other hand, coupled-cluster
theory is known to be the most elegant to effectuate the electron correlation \cite{2,3}. It is,
therefore, the combination of these two methods will surely be the solution to the problem.

The first step put forward in this direction is by Kaldor and coworkers.
They implemented relativistic counterpart of the Fock-space multi-reference
coupled cluster theory (FSMRCC) for the cause and applied extensively to both
atomic and molecular systems \cite{4,5,6,7,8}.
The original idea of FSMRCC theory is based on the construction of
an effective Hamiltonian using Bloch-Lindgren equation
to extricate some of the roots of the Hamiltonian matrix from the set of
entire eigen spectrum \cite{9a,9b,9c,9d,9,10,10a}.
The effective Hamiltonian variant of FSMRCC theory works within a reduced dimensional space,
called model space, is the part of the correlation space chosen to construct the effective
Hamiltonian and rest of the space is known as orthogonal space. The linear combination of
suitably chosen active configurations based on energetic criteria is used to construct the model space.
An operator, called wave operator is defined through which the contribution of orthogonal space included, is
the tool to construct the effective Hamiltonian. 
Finally, the diagonalization of the effective Hamiltonian matrix includes the correlation
contribution of the model space and results into the set of desired eigenvalues.
The problem associated with a reduced dimension effective Hamiltonian is the problem of intruder state,
which leads to the failure in convergence.
It appears that the effective Hamiltonian formalism of the FSMRCC theory is not straight forward,
conceptually difficult and lot of complicacy is associated with it. 

An alternative elegant approach to tackle the problem
is to use of the equation-of-motion variant of the coupled cluster theory (EOMCC) \cite{11,12,13,14,15,15a}.
The EOMCC is operationally a two step process: (i) solution of the coupled cluster equation for the N electron
determinant and (ii) construction and diagonalization of the effective Hamiltonian matrix
in the (N-1) electron determinantal space to get the desired set of eigenvectors and eigenvalues.
As EOMCC simultaneously treats two Hilbert space (N and N-1 electron space) in a single problem, the
effect of relaxation is also taken care which plays a key role
in the accurate description of the electronic states.
The dynamic part of the electron correlation is taken care by the exponential structure of the CC operator whereas
non-dynamical part comes through the diagonalization of the effective Hamiltonian in the configuration space.
We must admit that the EOMCC for the single ionization problem is equivalent to the (0,1) sector
FSMRCC theory and produces identical results for the principal peaks \cite{16,17}.
The superiority of the EOMCC method over FSMRCC theory is that the numerical instability
due to the problem of intruder states in FSMRCC does not arise in EOMCC as it is an eigenvalue problem.
The EOMCC is capable of giving shake-up states, which play important role in explaining various photo-ionization spectra \cite{17e}.

The EOMCC can be viewed as a multi-state approach where multiple states are obtained in a single 
calculation and are treated on equal footing. It works
within a single reference description to describe the complex multi-configurational
wave function. It is pertinent to say that EOMCC behaves properly at the non-interacting limit
but not rigorously size extensive \cite{17a,17b}. The error due to the size extensivity is reduced
due to the presence of higher order block. Furthermore, the eigenstates in EOMCC method are obtained directly contrary to the
propagator approaches though both the methods are of EOM structure \cite{17c,17d}.     
 
Recently, we stepped into the domain of fully 4-
component relativistic EOMCC and
employed to calculate single ionization
and double ionization potentials but that was for the atomic systems in their closed-shell configuration \cite{18,19,20}.
The molecular relativistic calculations are always more tedious than the atomic ones.
The spherical symmetry can be exploited in the atomic case, which allows the separation of 
the radial and angular part to use of the reduced matrix elements. The evaluation of radial integrals
can be done using the numerical integration. It reduces the
computational scaling. However, the method is less straight forward as each of the radial integral has
to be multiplied by the corresponding angular factor. The use of anti-symmetrized quantities (two-body matrix elements)
are common in molecular calculations, which is not suitable for the spherical implementation in the atomic case as different angular factor
will arise for the direct and exchange part of the radial integrals. The complexity associated with the
atomic calculations is more than compensated by the need to solve only for the radial equations. This
allows the use of very large basis set and to correlate more number of electrons with numerically
evaluated radial integrals to achieve better accuracy of results.

The relativistic effective core potential (RECP) is routinely used in molecular relativistic calculations \cite{20a}.
In RECP, only valence and some of the outer-core electrons are treated explicitly and rest of the electrons are
replaced by an effective RECP operator. This includes the simulating interaction of explicitly treated
electrons with those, which are excluded from the RECP calculation. There are variety of RECPs depending
on how the RECPs are optimized \cite{20b}. Among the various RECPs, the RECP with spin-orbit (SO) interaction is the
most popular one. This is generally done on the basis of separation of the electrons into core and
valence and between the scalar and spin-orbit relativistic effects according to the energy.
It allows exclusion of a large number of chemically inert electrons from the SCF calculations,
which eventually reduces the computational costs for the correlation calculation as compared
to the all electron 2-component and 4-component calculation.
The problem associated with this approach is the lack of control 
over accuracy.

Hirata was first to implement relativistic EOMCC for the purpose of molecular calculations \cite{21}.
He combined different electron correlation methods, basis set, and relativistic treatment to make 
a composite method.
The dynamic part of the electron correlation is taken care with a low rank method including the scalar relativistic
effect and employed various basis sets to enable complete basis set extrapolation. The non dynamical correlation 
is treated using EOMCC method with small basis set. Finally, the SO effect is added as the energy difference between
the RECP+SO with RECP calculated using a low rank correlated method. This approach cancels some of the
errors associated with the RECP methods. 
We would rather call Hirata's treatment as a good compromise of the different many-body effects to get reasonable results.
This approach does not address the complex interplay between the relativistic and correlation effects, which has been      
taken into account using 4-component single particle wave function
and the Dirac-Coulomb Hamiltonian along with the correlation treatment is done by the EOM-CCSD method.

In this article, we consider the implementation of fully 4-component relativistic EOMCC method 
to calculate ionization potentials of molecular systems within the single- and double- excitation
approximation (EOM-CCSD method). Pilot calculations of molecular ionization potential using EOM-CCSD method
are presented. We have also presented results by constructing the ground state wave function at the first order
perturbation theory level, which corresponds to the second order perturbation energy as the ground
state energy. We call this as EOM-MBPT(2). These results are compared with the EOM-CCSD results
to understand the role of electron correlation. 
To justify the fact that the relativistic and electron correlation effects are non-additive, we
have chosen HF as an example system. Both exact 2-component (X2C) and 4-component EOMCC calculations
are performed on it.

The manuscript is organized as follows. A brief description of the EOMCC
theory in the context of ionization potential is given in Sec. \ref{sec2}
and the computational details are presented in Sec. \ref{sec3}.
We have presented results and discuss on them in Sec. \ref{sec4} before
making our final remarks in Sec. \ref{sec5}.
We are consistent with atomic unit unless stated.
\begin{table}[t]
\caption{Comparison of correlation energy from MBPT(2) (E$_{corr}^{(2)}$) and CCSD (E$_{corr}^{(ccsd)}$) of HF as a fuction of number of basis functions.}
\begin{ruledtabular}
\begin{tabular}{lrrrrrrrrrr}
No of orbitals & \multicolumn{2}{c}{X2C-EOMCC} & \multicolumn{2}{c}{4C-EOMCC} \\
\cline{2-3} \cline{4-5}\\
& E$_{corr}^{(2)}$ &E$_{corr}^{(ccsd)}$& E$_{corr}^{(2)}$ & E$_{corr}^{(ccsd)}$\\  
\hline
& & \\
200  &\,-0.2921&\,-0.2930&\,-0.2921&\,-0.2929\\
220  &\,-0.3176&\,-0.3167&\,-0.3175&\,-0.3166\\
250  &\,-0.3482&\,-0.3477&\,-0.3480&\,-0.3475\\
280  &\,-0.3619&\,-0.3615&\,-0.3615&\,-0.3612\\
308  &\,-0.3659&\,-0.3655&\,-0.3655&\,-0.3651\\
\end{tabular}
\end{ruledtabular}
\label{x2ccorr}
\end{table}
\begin{table*}[ht]
\caption{Variation of ionization potential (in eV) as a function of basis function of $\mathrm{HF}$ molecule}
\begin{ruledtabular}
\begin{tabular}{lrrrrrrrrrr}
No of orbitals & \multicolumn{5}{c}{X2C-EOMCC} & \multicolumn{5}{c}{4C-EOMCC} \\
\cline{2-6} \cline{7-11}\\
& ${5\Pi}$  & ${4\Pi}$ &${3\Pi}$& ${2\Sigma}$& ${1\Sigma}$& ${5\Pi}$  
& ${4\Pi}$  &${3\Pi}$  &${2\Sigma}$&${1\Sigma}$\\
\hline
& & \\
${200}$&\,16.0432&\,16.0862&\,19.9960&\,39.4931&\,696.7720&\,16.0433&\,16.0859&\,19.9961&\,39.4968&\,696.8845\\
${220}$&\,16.1150&\,16.1549&\,20.0505&\,39.5434&\,696.1626&\,16.1150&\,16.1546&\,20.0506&\,39.5472&\,696.2763\\
${250}$&\,16.1366&\,16.1780&\,20.0648&\,39.5638&\,696.1292&\,16.1365&\,16.1777&\,20.0649&\,39.5675&\,696.2421\\
${280}$&\,16.1400&\,16.1800&\,20.0677&\,39.5727&\,696.3374&\,16.1399&\,16.1798&\,20.0677&\,39.5763&\,696.4479\\
${308}$&\,16.1398&\,16.1798&\,20.0681&\,39.5765&\,696.4326&\,16.1397&\,16.1796&\,20.0682&\,39.5800&\,696.5410\\
\end{tabular}
\end{ruledtabular}
\label{ipx2c}
\end{table*}

\section{theory}\label{sec2}
The starting point for the EOMCC calculation for the ionization problem is the 
solution of the reference wave function, which is the N electron CC closed-shell ground state wave function.
The ground state wave function in the CC method is defined as 
\begin{equation}
{|\Psi_{0}\rangle=e^{\hat T}|\Phi_{0}\rangle}.
\end{equation}
Where, ${|\Phi_{0}\rangle}$ is the single slater determinant  
which is the closed-shell $N$ electron Dirac-Hartree-Fock reference
determinant in our case.
 $\hat T$ is the cluster operator, within the CCSD approximation is represented as 
\begin{equation}
\begin{split}
T=& T_{1}+T_{2}\\
             =&  \sum\limits_{{i,a}} t_{i}^{a}a_{a}^{\dag}a_{i} +
     \sum\limits_{\stackrel{a<b}{i<j}}t_{ij}^{ab}a_{a}^{\dag}a_{b}^{\dag}a_{j}a_{i} .
\end{split}
\end{equation}
The $i,j (a,b)$ are the indices for the occupied (virtual) spinors. 
The cluster operators are solved by the following simultaneous non-linear algebraic equations.
\begin{equation}
{\langle \Phi_{i}^{a}|e^{-T}\hat He^{T}|\Phi_{0}\rangle=0,\,\,\, \langle \Phi_{ij}^{ab}|e^{-T}\hat He^{T}|\Phi_{0}\rangle=0 }.
\end{equation}
where, $|\Phi_{i}^{a}\rangle$ and $|\Phi_{ij}^{ab}\rangle$ are the singly and doubly excited determinant with reference 
to the $N$ electron closed-shell Dirac-Hartree-Fock determinant.
Finally, the ground state energy is obtained by solving the 
equation for the energy,
\begin{equation}
{E_{ccsd}=\langle \Phi_{0}|e^{-T}\hat He^{T}|\Phi_{0}\rangle} , 
\end{equation}
where, $\hat H$ is the Dirac-Coulomb Hamiltonian which is, 
\begin{eqnarray}
{\hat H_{DC}} &=&\sum_{A}\sum_{i} [c (\vec {\alpha}\cdot \vec {p})_i + \beta_i m_0c^{2} + V_{iA}] \nonumber\\
       &+& \sum_{i>j} \frac{1}{r_{ij}} {\mathbb{1}_4}.
\end{eqnarray}
The $\alpha$ and $\beta$ are the usual Dirac matrices. $V_{iA}$ stands for potential energy operator
for the $i^{th}$ electron in the field of nucleus A.
$m_0c^2$ is the free electron rest mass energy where c is the speed of light.

\begin{figure}[h]
\centering
\begin{center}
\includegraphics[height=1.7 cm, width=8.2 cm]{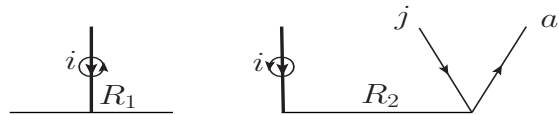}
\caption { Diagrammatic representation of $R_1$ and $R_2$ operator.}
\label{fig1}
\end{center}  
\end{figure}

In the EOMCC approach for the single electron ionization problem, the wave function for the $k^{th}$ target state is created
by the action of a linear operator ($R(k)$) on the single reference coupled cluster wave function $|\Psi_{0}\rangle$, 
\begin{equation}
{|\Psi_k\rangle=R(k)|\Psi_0\rangle}, 
\end{equation}
Within the CCSD approximation $R(k)$ is also approximated to 
\begin{equation}
\begin{split}
R(k) =& R_{1}+R_{2}\\
             =&  \sum_{i} r_{i} a_i +  \sum\limits_{\stackrel{a}{i < j}} r_{i j}^{a} a^{\dag}_a a_j a_i ,
\end{split}
\end{equation}
The diagrammatic representation of the $R_1$ and $R_2$ operator is presented in Fig. \ref{fig1}
and are one rank higher than the CC operators.

\begin{figure}[h]
\centering
\begin{center}
\includegraphics[height=4.0 cm, width=7.0 cm]{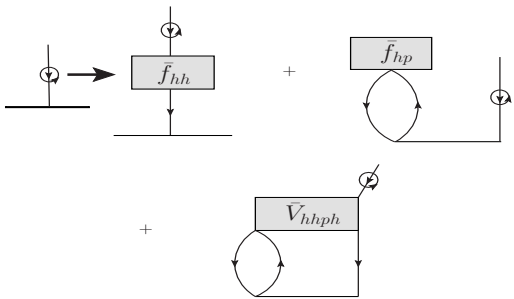}

\caption {Diagrams contributing to the 1h block.}
\label{r1block}
\end{center}  
\end{figure}

\begin{figure}[h]
\centering
\begin{center}
\includegraphics[height=6.0 cm, width=8.0 cm]{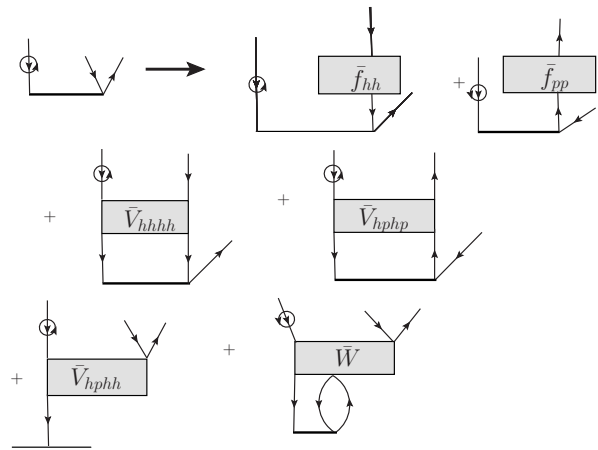}
\caption{Diagrams contributing to the 2h-1p block.}
\label{r2block}
\end{center}
\end{figure}

The energy of the $k^{th}$ ionized state is determined by the equation
\begin{eqnarray}
\begin{split}
e^{-T}HRe^{T}|\Phi_{0}\rangle= e^{-T}He^{T}R|\Phi_{0}\rangle=&\bar {H}R|\Phi_{0}\rangle\\
                                                             =& ER|\Phi_{0}\rangle,
\end{split}
\label{him}
\end{eqnarray}
It is assumed that the $e^{T}$ and $R$ commute as they are the strings of same quasi-particle
creation operator. Here, $\bar {H}=e^{-T}He^{T}$ is the effective Hamiltonian and E is the
energy of the ionized state, is the sum of the $E_{ccsd}$ and the corresponding ionization
potential. Subtraction of $E_{ccsd}$ from the eq. \ref{him}  takes the form as
\begin{equation}
[\bar{H},\hat R(k)]|\Phi_{0} \rangle=\Delta E_k \hat R{(k)}|\Phi_{0} \rangle, \ \ \ \ \forall k, 
\end{equation}
That is why this approach is called EOMCC in analogy to the Heisenberg's equation of motion for the
excitation operator $R(k)$.
A correlated determinantal space of  $|\phi_{i}\rangle$ and $|\phi_{i j}^{a}\rangle$ (1h and 2h-1p)
with respect to $|\Phi_{0}\rangle$ is chosen to project the above equation to get the desired ionization
potential values, $ \Delta E_k$.  
\begin{equation}
{\langle \phi_{i}[\bar{H}, R(k)]|\phi_{0}\rangle=\Delta E_k R_i},
\end{equation}
\begin{equation}
{\langle \phi_{ij}^{a}[\bar {H},R(k)]|\phi_{0} \rangle =\Delta E_k R_{ij}^{a}},
\end{equation}
The above equations can be represented in the matrix form as 
${\bar{H} R=R\Delta E_k}$.
The anti-symmetrized diagrams contributing to the 1h and 2h-1p block are presented in the Fig. \ref{r1block}
and Fig. \ref{r2block} respectively.
The evaluation of these diagrams is done by constructing one-body, two-body and three-body 
intermediate diagrams.
This requires the solution of the coupled cluster ground state amplitude equations.
With the converged $T_1$ and $T_2$ amplitudes from CC ground state calculations, these intermediate diagrams are constructed
by contracting one-body and two-body parts of the effective Hamiltonian matrix elements.
There are three distinct type one-body, four two-body and one three-body intermediate diagrams
are required for the calculation of ionization potential using EOM-CCSD method.
We call these as 
$\bar f_{hh}$, $\bar f_{pp}$, $\bar f_{hp}$,  $\bar v_{hphh}$, $\bar v_{hhhh}$, $\bar v_{hhph}$, $\bar v_{hhph}$, and $\bar W$.
Here $\bar f$, $\bar v$ and $\bar W$ stand for one-body, two-body and three-body intermediates respectively.
The algebraic expression as well as diagrammatic of the intermediate diagrams can be found in \cite{rod}.
All these intermediate diagrams are inserted in between the diagrams contributing to the 1h and 2h-1p block.
A circled arrow represents a detached occupied orbital.
\begin{table}
\caption{Bond length(in $A^{0}$), SCF energy (${E_{DF}^{0}}$), correlation energies from the MBPT(2) (${E_{corr}^{(2)}}$) and
CCSD (${E_{corr}^{(ccsd)}}$) methods for different systems.}
\begin{ruledtabular}
\begin{tabular}{l r r r r }
Molecule &  Bond Length\cite{27} & {${E_{DF}^{0}}$} &  {${E_{corr}^{(2)}}$}& {${E_{corr}^{(ccsd)}}$}  \\
\hline
HF &\,0.9168&\,-100.1604 &\,-0.3649&\,-0.3646\\
HCl&\,1.2750&\,-461.5644 &\,-0.6228 &\,-0.6382\\
HBr&\,1.4140&\,-2605.6330&\,-1.6500&\,-1.5873\\
HI&\,1.6090&\,-7116.3860&\,-2.0049&\,-1.9134\\
Cl$_2$&\,1.9870&\,-921.9144&\,-1.2180&\,-1.2404\\
Br$_2$&\,2.2810&\,-5210.0830&\,-2.9822&\,-2.8543\\
\end{tabular}
\end{ruledtabular}
\label{tab1}
\end{table}

The dimension of the $\bar{H}$ matrix is quite large $(nh+nh^2np,nh+nh^2np)$ for the relativistic
calculations in a reasonable basis. It is, therefore, following a full diagonalization
scheme is not at all a good idea. Here $nh$ and $np$ represents the number of hole and particle respectively.
The Davidson algorithm \cite{davidson}, which is an iterative diagonalization scheme is implemented 
for the diagonalization purpose, $\bar{H}$ to get the desired set of 
eigenvalues, $\Delta E_{k}$ and the corresponding eigenvectors.
It avoids computation, storage and diagonalization of the full matrix.
The EOMCC can be regarded as the diagonalization of the coupled cluster similarity transformed
Hamiltonian in the configuration space. This makes the EOMCC as a hybrid method of coupled cluster and configuration interaction (CI).
\begin{table*}
\caption{Vertical IPs (in ev) of  ($XH(X=F, Cl, Br, I)$), $Cl_2$ and $Br_2$  using EOM-CCSD methods.}
\begin{ruledtabular}
\begin{tabular}{ l r r r r}
Molecule  & Ionizing State & EOM-MBPT(2)& EOM-CCSD  & Experiment\\
\hline
HF        &\,${5\,\Pi}$&\,16.1709&\,16.1380  &\, 16.1200 \cite{28} \\
          &\,${4\,\Pi}$&\,16.2109&\,16.1777  &\,                 \\
          &\,$3\,\Sigma$&\,20.0648&\,20.0667  &\,19.8900 \cite{28}\\
          &\,$2\,\Sigma$&\,39.5239&\,39.5802  &\,39.6500 \cite{28}\\
          &\ $1\,\Sigma$&\,697.0884&\,696.6777 &\,694.0000\cite{28}\\
HCl       &\,$9\,\Pi$&\, 12.8248     &\,12.8079  &\,12.7450 \cite{29a} \\
          &\,$8\,\Pi$&\, 12.9090     &\,12.8917  &\,12.8300\cite{29a}\\
          &\,$7\,\Sigma$&\, 16.8321     &\,16.8230  &\,          \\
          &\,$6\,\Sigma$&\, 25.8646     &\,25.8799  &\,           \\
HBr       &\,$18\,\Pi$&\,11.8294&\,11.6977   &\,11.6450 \cite{30a}\\
          &\,$17\,\Pi$&\,12.1693&\,12.0343   &\,11.9800\cite{30a}\\
          &\,$16\,\Sigma$&\,15.9093&\,15.8169&\,15.6500\cite{30a}\\
HI       &\,$27\,\Pi$&\,10.6763&\, 10.4229  &\,10.3880 \cite{30b}\\
         &\,$26\,\Pi$&\,11.3628&\, 11.0998  &\,11.047\cite{30b}\\
Cl$_2$     &\,$17\,\Pi$&\,11.6842  &\,11.6679&\, 11.5900 \cite{30c}\\
           &\,$16\,\Pi$&\,11.7774  &\,11.7604&\,         \\
           &\,$15\,\Pi$&\,14.6353  &\,14.4969&\, 14.4000 \cite{30c}\\
           &\,$14\,\Pi$&\,14.7138  &\,14.5751&\,          \\
Br$_2$     &\,$35\,\Pi$&\,10.5681&\,10.4370&\,  10.5180 \cite{31} \\
           &\,$34\,\Pi$&\,10.9252&\,10.7897&\,  10.8670  \cite{31}\\
\end{tabular}
\end{ruledtabular}
\label{tab2}
\end{table*}
\section{computational details}\label{sec3}
The one-body and two-body matrix elements are generated with the help
of DIRAC10 program package \cite{22}. The finite atomic orbital basis consists of scalar, real Gaussian functions.  
The large components of the basis set are contracted and the small components are in uncontracted except
for the Br$_2$ molecule where both the large and small component are uncontracted in nature.
The small component of the basis set is generated by imposing restricted kinetic balance (RKB)
condition with the large components. This RKB is done by the preprojection in scalar basis and the
unphysical solutions are removed by diagonalizing the free particle Hamiltonian. The DIRAC10 uses
Gaussian charge distribution for the nuclear potential. The nuclear parameters used in our calculations
are all default values. We adopted cc-pVQZ basis set \cite{23} for H atom in all the calculations. In the calculation
of HF and HCl molecule, the basis set chosen both for F and Cl atom is aug-cc-pCVQZ \cite{24,25}. The dyall.acv4z \cite{26} basis
is chosen for Br and I for the calculations of HBr and HI.  The basis set chosen for Cl and Br atom are aug-cc-pCVQZ \cite{25}
and dyall.cv3z \cite{26} respectively in the calculation of  Cl$_2$ and  Br$_2$. 
We have taken into account C$_{2v}$ symmetry to generate the single particle orbitals and two body matrix elements in all the calculations
and none of the electrons are frozen for the correlation calculations.
In the implemented version of X2C SCF in DIRAC10, the large component of the basis is uncontracted in nature.
Therefore, the 4-component calculations to compare with X2C-EOMCC, both the large and small component of the basis are also taken as uncontracted 
fashion to generate the same determinantal space.
The matrix elements of the intermediate diagrams are stored putting a cutoff  of $10^{-12}$ to save 
storage requirement as contribution of the matrix elements beyond 12-decimal places is very less.
To debug our newly implemented relativistic EOM-CCSD code, we benchmarked our results  with the Fock-space
MRCC code of DIRAC10 for the ionization problem with same basis, same convergence criteria and equal number of DIIS 
space as these two methods in principle are supposed to produce identical results. We have achieved identical results
for MBPT(2) correlation energy, 10-digit agreement for the CCSD correlation energy and
8-digit agreement for ionization potential values.
This agreement is achieved independent of the choice of molecules as well as of the basis sets.
The discrepancy beyond this limit could be due to the different 
convergence algorithm and the use of cutoff in the construction of the intermediate diagrams.
The experimental bond length used in our calculations are taken from the ref \cite{27}.
In our calculations we have used $10^{-6}$ as convergence cutoff for the Davidson algorithm and $10^{-10}$
for the ground state coupled cluster equations.
The numerical labeling of the ionized states are done from the inner to the outer.
\section{Results and Discussion}\label{sec4}
We present numerical results of our calculations using 4-component EOM-CCSD method developed by us for the
calculation of ionization potentials of molecular systems by removing an electron from their
closed-shell configuration. We also present results using an intermediate scheme, 
EOM-MBPT(2) which uses first order perturbed wave function for the construction of ground state wave function.
We applied both these methods to HF, HCl, HBr, HI, Cl$_2$ and Br$_2$ molecule.
Comparison has been done between the X2C-EOMCC and 4-component EOMCC to justify the fact that
the relativistic and correlation effects are non-additive in nature taking example of HF molecule.

In table \ref{x2ccorr}, we present the correlation energies from MBPT2 ($E^{(2)}_{corr}$) and CCSD ($E^{(ccsd)}_{corr}$)
method as a function of number of basis functions for both the X2C-EOMCC and 4-component EOMCC of HF molecule.
The SCF energy for the 4-component calculation is -100.161280, whereas
it is -100.156272 for the X2C calculations. In both the calculations the basis functions are used in uncontracted fashion
. We have started our calculation with 200 active orbitals for the calculation of correlation energies using MBPT2 and CCSD
and keep on increasing upto 308 which is the maximum number of orbitals possible to generate for the opted basis.
In correlation calculation the correlation space is identical for both the X2C and 4-component calculations,
therefore, in principle correlation energy must be the same but the values obtained are not identical.
The difference in the  SCF energy is in the order of 0.01 au. The same difference should reflect in
the correlation energy calculations if these two effects are additive. 
The outcome is clearly because of the non-additivity of the relativity and electron correlation.
The difference between the two scheme is less for the calculation using 200 active orbitals and increases
with increase in the correlation space. The deviation between the X2C-EOMCC and 4-component EOMCC calculations
is expected to be more for the molecules containing heavier atoms as effect of relativity is dominant factor for the heavy atoms. 

In table \ref{ipx2c},  the results of variation of ionization potential as a function
of basis function both for the X2C-EOMCC and 4-component EOMCC of $HF$ molecule with different number of active orbitals are presented.
The difference between the X2C-EOMCC and 4-component EOMCC is negligible for the outer orbitals but
the more for the inner orbitals.
The deviation is in the fourth digit for the valence orbitals whereas the difference is in the 
first digit for the inner core orbitals after the decimal.
The deviation increases towards the core orbitals as the effect of relativity increases. 
It is expected that the difference will be more for the
inner orbitals as the effects of relativity is dominant near the nucleus.
The effect will be prominent for the heavier systems as effects of relativity plays more decisive role in those systems. 
The results further justifies the argument of non-additivity of relativity and electron correlation.
\begin{figure}
\centering
\begin{center}
\includegraphics[height=4.8cm, width=7.5 cm]{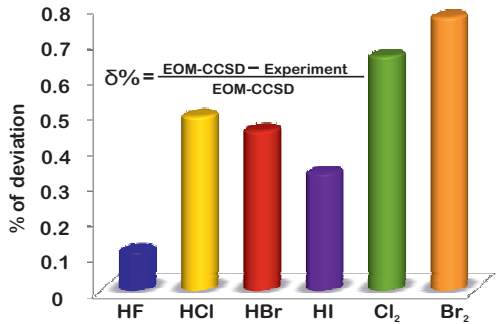}
\caption {(color online)Relative deviations in $(\%)$ from the experimental values.}
\label{fig2}
\end{center}  
\end{figure}

In table \ref{tab1}, we present the equilibrium bond length used in the calculations of considered diatomic molecules
and also the SCF energy($E^0_{DF}$), correlation energy from the 2-nd order perturbation theory
($E^{(2)}_{corr}$) and CCSD method ($E^{(ccsd)}_{corr}$). The reported SCF results are calculated using the DIRAC10
and correlation calculations are done with relativistic code developed by us for the purpose of ground state energy calculations within
the single- and double- excitation approximation. 

In table \ref{tab2}, we present results of vertical ionization potential of diatomic molecules using EOM-CCSD
and EOM-MBPT(2) method.
The results of our calculation of ionization potentials are compared with
the available experimental values.
Our EOM-CCSD results for the valence orbitals show good agreement with the experimental values and the difference is less than
0.1 eV. The difference is slightly more for the inner orbitals which  
is expected that the extent of accuracy will definitely be less as compared to 
the valence orbitals as we have used Dirac-Coulomb Hamiltonian in our calculations.
The higher order relativistic effects specially the Breit interactions for the neutral molecules
have sigificant contribution for the inner orbitals. On the other hand, the deviation for the EOM-MBPT(2) 
is more as a dominant part of dynamic correlation is missing in the scheme. 
We present the deviation of valence ionization
calculations as $\delta \%$ in Fig \ref{fig2}. In all the calculated systems we have achieved an accuracy
of less than 1$\%$ with the standard values. The maximum deviation is for the Br$_2$ molecule 
which is 0.77$\%$ and least for HF is 0.11$\%$. 
One possible reason for the deviation in Br$_2$ molecule could be the basis employed is not adequate
for the exact description of the ionized states.
  
\section{conclusion}\label{sec5}
We have successfully implemented 4-component relativistic equation-of-motion coupled cluster method (EOM-CCSD) 
to calculate ionization potential of molecular system in their closed-shell configuration.
We presented numerical results of our calculation using both EOM-CCSD and EOM-MBPT(2) method.
Our results are found to be in excellent agreement with the experimental values.

\section*{acknowledgments}
H.P., S.S., N.V. and S.P. acknowledge the grant from CSIR XIIth five year plan project on Multi-scale Simulations of 
Material (MSM) and facilities of the Center of Excellence in Scientific Computing at CSIR-NCL.
H.P and S.S acknowledge the 
Council of Scientific and Industrial Research (CSIR) for their fellowship. SP acknowledges grant from DST, 
J. C. Bose fellowship project towards completion of the work.

\end{document}